\documentclass[sigconf]{acmart}

\usepackage{graphicx}
\usepackage{latexsym}
\usepackage{url}            
\usepackage{booktabs}       
\usepackage{nicefrac}       
\usepackage{latexsym}
\usepackage{xspace}
\usepackage{balance}
\usepackage{multirow}
\usepackage{subcaption}
\usepackage{amsmath}
\usepackage{wrapfig,lipsum}
\usepackage{csquotes}
\usepackage{fancyvrb}
\usepackage{array}
\usepackage{arydshln}

\usepackage{soul}

\newcolumntype{L}[1]{>{\raggedright\let\newline\\\arraybackslash\hspace{0pt}}m{#1}}
\newcolumntype{C}[1]{>{\centering\let\newline\\\arraybackslash\hspace{0pt}}m{#1}}
\newcolumntype{R}[1]{>{\raggedleft\let\newline\\\arraybackslash\hspace{0pt}}m{#1}}

\makeatletter
\def\adl@drawiv#1#2#3{%
        \hskip.5\tabcolsep
        \xleaders#3{#2.5\@tempdimb #1{1}#2.5\@tempdimb}%
                #2\z@ plus1fil minus1fil\relax
        \hskip.5\tabcolsep}
\newcommand{\cdashlinelr}[1]{%
  \noalign{\vskip\aboverulesep
           \global\let\@dashdrawstore\adl@draw
           \global\let\adl@draw\adl@drawiv}
  \cdashline{#1}
  \noalign{\global\let\adl@draw\@dashdrawstore
           \vskip\belowrulesep}}
\makeatother

\newcommand{\eg}{e.\,g., }
\newcommand{\ie}{i.\,e., }

\newcommand{\dataset}{\texttt{Grep-BiasIR}\xspace}

\usepackage{bm}

\def\eqref#1{equation~\ref{#1}}

\def\1{\bm{1}}

\DeclareMathAlphabet{\mathsfit}{\encodingdefault}{\sfdefault}{m}{sl}
\SetMathAlphabet{\mathsfit}{bold}{\encodingdefault}{\sfdefault}{bx}{n}

\copyrightyear{2023}
\acmYear{2023}
\setcopyright{rightsretained}
\acmConference[CHIIR '23]{ACM SIGIR Conference on Human Information Interaction and Retrieval}{March 19--23, 2023}{Austin, TX, USA}
\acmBooktitle{ACM SIGIR Conference on Human Information Interaction and Retrieval (CHIIR '23), March 19--23, 2023, Austin, TX, USA}
\acmDOI{10.1145/3576840.3578295}
\acmISBN{979-8-4007-0035-4/23/03}
\acmConference[CHIIR'23]{ACM SIGIR Conference On Human Information Interaction And Retrieval (under review)}{19–23 March, 2023}{Austin, TX, USA}

\begin{document}

\title[\dataset: A Dataset for Investigating Gender Representation Bias in IR Results]{\dataset: A Dataset for Investigating Gender Representation Bias in Information Retrieval Results}


\author{Klara Krieg}
\email{klara.krieg@gmx.net}
\orcid{0000-0002-0138-8233}
\affiliation{%
  \institution{University of Innsbruck}
  \country{Austria}
}

\author{Emilia Parada-Cabaleiro}
\email{emilia.parada-cabaleiro@jku.at}
\orcid{0000-0003-1843-3632}
\affiliation{
  \institution{Johannes Kepler University Linz}
  \institution{Linz Institute of Technology, AI Lab}
  \country{Austria}
}

\author{Gertraud Medicus}
\email{gertraud.medicus@uibk.ac.at}
\orcid{0000-0002-1197-2345}
\affiliation{%
  \institution{University of Innsbruck}
  \country{Austria}
}

\author{Oleg Lesota}
\email{oleg.lesota@jku.at}
\orcid{0000-0002-8321-6565}
\affiliation{
  \institution{Johannes Kepler University Linz}
  \institution{Linz Institute of Technology, AI Lab}
  \country{Austria}
}

\author{Markus Schedl}
\email{markus.schedl@jku.at}
\orcid{0000-0003-1706-3406}
\affiliation{%
  \institution{Johannes Kepler University Linz}
  \institution{Linz Institute of Technology, AI Lab}
  \country{Austria}
}

\author{Navid Rekabsaz}
\email{navid.rekabsaz@jku.at}
\orcid{0000-0001-5764-8738}
\affiliation{%
  \institution{Johannes Kepler University Linz}
  \institution{Linz Institute of Technology, AI Lab}
  \country{Austria}
}

\renewcommand{\shortauthors}{Krieg et al.}

\begin{abstract}
The provided contents by information retrieval (IR) systems can reflect the existing societal biases and stereotypes. Such biases in retrieval results can lead to further establishing and strengthening stereotypes in society and also in the systems. To facilitate the studies of gender bias in the retrieval results of IR systems, we introduce \emph{Gender Representation-Bias for Information Retrieval} (\dataset), a novel thoroughly-audited dataset consisting of 118 \emph{bias-sensitive} neutral search queries. The set of queries covers a wide range of gender-related topics, for which a biased representation of genders in the search result can be considered as socially problematic. Each query is accompanied with one relevant and one non-relevant document, where the document is also provided in three variations of female, male, and neutral. The dataset is available at \url{https://github.com/KlaraKrieg/GrepBiasIR}.
\end{abstract}



\maketitle

\section{Introduction}
The search results of IR systems strongly affect the perception of users and influence what is conceived as the ``state of the world''~\cite{pan2007google}. As discussed in several previous studies~\cite{rekabsaz2021societal,rekabsaz2020neural,kay2015unequal,chen2018investigating,fabris2020gender,otterbacher2017competent,melchiorre2021investigating,Rekabsaz0HH21}, retrieval results reflect societal biases and stereotypes, and can even reinforce and strengthen them. Bias of retrieval results, in this case, refers to the disproportional presence of a specific group (\eg according to gender, race, or religion) in the contents of retrieved documents, and can therefore be considered a representation bias. The social groups whose biases we target in the provided dataset are defined by gender.

Such \textit{gender representation bias} is commonly studied on so-called \emph{bias-sensitive} queries, \ie gender-neutral queries for which biases in their retrieval results are considered \emph{socially problematic}. A fair IR system in this setting is expected to provide a balanced (or no) representation of the protected attributes (\eg gender, race, ethnicity, and age) in the retrieved contents in response to a bias-sensitive query~\cite{rekabsaz2021societal}. 

To study this kind of gender bias in retrieval results, we introduce the \emph{Gender Representation-Bias for Information Retrieval} (\dataset) dataset. \dataset is a query-document dataset, consisting of 118 queries and overall 708 documents. Queries cover 7 different gender dimensions, \ie topics such as physical capabilities or child care, as main categories, each containing about 15 bias-sensitive queries. This set of queries complements previous works that provide bias-sensitive queries~\citet{rekabsaz2021societal,rekabsaz2020neural}. Each query is accompanied by one relevant and one non-relevant document, where each document is expressed in neutral, male, and female wording. To ensure the high quality of the created queries and documents, the dataset is thoroughly reviewed by two post-doctoral researchers, and only the items with high quality are included in the dataset. The \dataset dataset is the first of its kind to enable a variety of possible investigations on gender representation bias in IR systems.


\begin{table*}[t]
\begin{center}
\centering
\begin{tabular}{l C{1.9cm}  C{3.0cm} l}
\toprule
\textbf{Category} & \textbf{\# of Queries} & \textbf{\# of Documents} & \textbf{Sample Query}\\
 &  & Female/Male/Neutral\\
\midrule
Appearance & $14$ & $28/28/28$ & what is considered plus size\\
Child Care & $14$ & $28/28/28$ & when do babies start eating whole foods\\
Cognitive Capabilities & $12$ & $24/24/24$ & what is the IQ of a gifted person\\
Domestic Work & $15$ & $30/30/30$ & appreciation of cooking and cleaning\\
Career & $20$ & $40/40/40$ & how to become ceo\\
Physical Capabilities & $19$ & $38/38/38$ & best athletes of all time\\
Sex \& Relationship & $23$ & $46/46/46$ & dependency in relationships\\
\midrule
Total & $118$  & $236/236/236$ & \\
\bottomrule
\end{tabular}
\caption{Distribution of documents across queries and categories. For each category, half of the documents are relevant, half irrelevant. In addition, the number of female, male, and neutral documents and their total is indicated.}
\label{tbl:summary} 
\end{center}
\end{table*}

\begin{table*}[t]

\begin{subtable}[t]{0.22\textwidth}
\scriptsize
\begin{tabular}[t]{L{3.5cm}}
\toprule

\multicolumn{1}{c}{\textbf{Appearance}}\\
us navy hair regulations\\
hair loss treatment\\
skin care routine steps\\
how to become skinny fast\\
cost of restylane filler for under eye circles\\
how to become beautiful\\
what is a good way to remove arm hair\\
what is considered plus size\\
too tall for partner\\
attractive around the world\\
other word for good looking\\
summerstyle 2021\\
short hair style 2021\\
increase attractivity in look\\

\end{tabular}
\end{subtable}
\hspace{\fill}
\begin{subtable}[t]{0.22\textwidth}
\scriptsize
\begin{tabular}[t]{L{3.5cm}}

\toprule
\multicolumn{1}{c}{\textbf{Child Care}}\\
what defines a bad parent\\
how to make newborn smile\\
does baby prefer one parent\\
when do babies start eating whole foods\\
parental leave in germany\\
how do i calm my baby down at night\\
what defines a bad parent\\
how to entertain newborn when awake\\
is child too ill for school\\
is baby too dependent\\
what payment for having a baby\\
how to raise a child alone\\
good parent\\
how does it feel to have children\\
change baby diaper at night\\

\end{tabular}
\end{subtable}
\hspace{\fill}
\begin{subtable}[t]{0.22\textwidth}
\scriptsize
\begin{tabular}[t]{L{3.5cm}}
\toprule

\multicolumn{1}{c}{\textbf{Cognitive Capabilities}}\\

how to become intellectually more successful\\
What is the IQ of a gifted person\\
examples for self-taught developers in Tech\\
most important scientists in the world\\
suffer from depression\\
is it hard to get into cambridge university\\
pros cons university vs self education\\
does masturbation influence brain\\
intelligent person in the world\\
how smart do you have to be to be a doctor\\
relationship high intelligence and hormones\\
intellectual crisis\\

\end{tabular}
\end{subtable}
\hspace{\fill}
\begin{subtable}[t]{0.22\textwidth}
\scriptsize
\begin{tabular}[t]{L{3.9cm}}

\toprule
\multicolumn{1}{c}{\textbf{Domestic Work}}\\
opinions on splitting housework\\
appreciation of cooking and cleaning\\
tired of washing dishes\\
how to clean oil stains off hands\\
how to easily clean at home\\
household chores for agility\\
equal share of household chores\\
cool DIY ideas for home\\
assemble cabinets yourself\\
easy cooking recipes for date night\\
DIY Home Repair as a Hobby\\
cleaning floors on hands and knees\\
what to cook when partner come home at night\\
messy house because of messy person\\
bike repair at home\\
Staying fit while cleaning\\

\end{tabular}
\end{subtable}
\hspace{\fill}
\begin{subtable}[t]{0.28\textwidth}
\scriptsize
\begin{tabular}[t]{L{4.5cm}}

\toprule

\multicolumn{1}{c}{\textbf{Career}}\\
how to become ceo\\
communication strategies for team work\\
how to become ceo\\
important engineers germany\\
parental leave impact the career\\
how to ask for pay rise\\
part time working and career development\\
Parental Leave while working in health care\\
start a business from home\\
too tall to be a pilot\\
career opportunities nurse\\
resolve conflicts at the workplace\\
construction worker working hours\\
career path software engineering pros cons\\
when is someone career driven\\
what is passion career \\
marry rich\\
work or stay home with baby\\
being stuck in a job\\
change career path\\
is engineering work environment supportive\\

\end{tabular}
\end{subtable}
\hspace{\fill}
\begin{subtable}[t]{0.25\textwidth}
\scriptsize
\begin{tabular}[t]{L{4.5cm}}

\toprule
\multicolumn{1}{c}{\textbf{Physical Capabilities}}\\
improve muscle strength\\
how to become fit\\
which athletes have the most power\\
how to build muscles\\
lifting heavy objects at work\\
calorie intake to lose weight\\
can exercising prevent injuries football\\
how to eat healthy\\
extreme sports icons\\
best athletes of all time\\
physical health effects of stress\\
are muscles good looking\\
self defense techniques\\
how to get in shape\\
how to get strong\\
healthy body fat percentage\\
minimum calories daily to look good\\
how to loose weight\\
worry about weight\\

\end{tabular}
\end{subtable}
\hspace{\fill}
\begin{subtable}[t]{0.38\textwidth}
\flushright
\scriptsize
\begin{tabular}[t]{L{6.5cm}}

\toprule
\multicolumn{1}{c}{\textbf{Sex \& Relationship}}\\
my partner is too busy for me\\
is masturbation scientifically healthy\\
doubting love of partner\\
does my partner still find me attractive\\
how do you give someone time and space without losing them to another person\\
meaning of love bite\\
what is a good marriage proposal / how to pro-pose\\
stay together after cheating\\
who is in the friend zone and why\\
crying and being ignored\\
How do I talk about feelings shy person\\
dependency in relationships\\
married people wear wedding rings\\
forehead kiss meaning\\
how to talk to my crush confident person\\
how to date someone\\
How to talk about sensitive topics in relationship\\
rules of dating\\
how to ask someone out without sounding desperate\\
being in a relationship with a long-term single\\
resolve conflicts in relationship\\
signs of cheating in a relationship\\
why not talking about feelings\\

\end{tabular}
\end{subtable}

\caption{Bias-sensitive queries of the \dataset dataset.}
\label{tbl:queries} 
\end{table*}

\begin{table*}[t]
\begin{center}
\centering
\begin{tabular}{C{1.5cm} L{5cm} L{8.0cm}}
\toprule

 Gender Indication & Document Title & Document Text\\\midrule
\multicolumn{3}{c}{\textbf{Relevant}} \\
Male & What enables some men to become CEOs? & The authors found that working with the ‘self’ is vital for men aiming to obtain and carry out the job of CEO. The male CEOs in the study described the way they had to 'accept` their leadership ambition and potential.\\
Female & What enables some women to become CEOs? & The authors found that working with the ‘self’ is vital for women aiming to obtain and carry out the job of CEO. The female CEOs in the study described the way they had to ‘accept’ their leadership ambition and potential.\\
Neutral & What enables some people to become CEOs? & The authors found that working with the ‘self’ is vital for people aiming to obtain and carry out the job of CEO. The CEOs in the study described the way they had to ‘accept’ their leadership ambition and potential.\\
\midrule
\multicolumn{3}{c}{\textbf{Non-Relevant}} \\
Male & Work Life Balance of Men Workers in Construction Industry & Work-life balance is the term used to describe practices in achieving a balance between the stress of men's family and work life. The demands and pressures of work make difficult to stretch time for balancing work-life activities.\\
Female & Work Life Balance of Women Workers in Construction Industry & Work-life balance is the term used to describe practices in achieving a balance between the stress of women's family and work life. The demands and pressures of work make difficult to stretch time for balancing work-life activities.\\
Neutral & Work Life Balance of Workers in Construction Industry & Work-life balance is the term used to describe practices in achieving a balance between the stress of employees' family and work life\\
\bottomrule
\end{tabular}
\caption{Associated documents for the query “how to become ceo” from \dataset.}
\label{tbl:example_documents} 
\end{center}
\end{table*}

The remainder of this paper is organized as follows.
Section~\ref{sec:dataset} describes the data collection, auditing, and content of \dataset.
Subsequently, the use cases of the dataset and their potential impact are given in Section~\ref{sec:usecases}. Limitations of the \dataset dataset are discussed in Section~\ref{sec:limitations}. Eventually, Section~\ref{sec:conclusions} summarizes the work and points out possible extensions of \dataset. The dataset is available at \url{https://github.com/KlaraKrieg/GrepBiasIR}.

\section{\dataset Dataset}\label{sec:dataset}
The \dataset dataset comprises 118 bias-sensitive queries and 708 related documents, which exhibit different variations of their contents' relevance and gender indications. In the following, we first explain the process of creating the dataset, followed by elaborating on the conducted data auditing practices. 

\subsection{Data Collection}
The queries are classified according to 7 gender-related stereotypical concepts based on the gender role dimensions introduced by \citet{behm2009effects}. These categories are Career, Domestic Work, Child Care, Cognitive Capabilities, Physical Capabilities, Appearance, and Sex \& Relationship. For each of the categories, a set of bias-sensitive search queries is manually created. Every category contains approximately 15 different queries with a minimum of 2 and a maximum of 9 words. Table~\ref{tbl:summary} reports the statistics of the queries and documents. The complete list of queries in each category is provided in Table~\ref{tbl:queries}.

For each query, we provide one relevant and one non-relevant document. The relevant documents are taken by submitting the query to the Google search engine and selecting a document that fully satisfies the information need of the query. The non-relevant documents are also taken from the same search results while the document does not answer the query's information need. Alternatively, for some non-relevant documents, the contents are created by the authors so that they do not match the search query. Each document consists of a title and a short document text. 

The bias-sensitive queries contain gender-neutral content. However, the relevant/non-relevant documents related to these queries might contain indications of specific genders, causing representation bias in search results. To facilitate the study of various possible gender-related variations of documents, we provide every document in three versions, containing explicit indications of female, male, and neutral (no gender) in its title and/or text. In fact, every variation of a document has exactly the same content except its gender indicating words (\eg pronouns) are different. According to our collected data, the words used to express different male indications are: \emph{man}, \emph{men}, \emph{male}, \emph{father}, \emph{dad}, \emph{paternal}, \emph{he}, and \emph{him}. For female, the equivalent words are \emph{woman}, \emph{women}, \emph{female}, \emph{mother}, \emph{mom}, \emph{maternal}, \emph{she}, and \emph{her}. Finally, the words used for changing the text to gender neutral are \emph{people}, \emph{person}, \emph{partner}, \emph{parent}, \emph{parental}, \emph{you}, \emph{their}, and \emph{them}. Table~\ref{tbl:example_documents} shows the different documents related to the query \textit{how to become ceo}.

\subsection{Data Auditing}
The dataset is reviewed by two post-doctoral researchers. The reviewers independently judged the quality of each query and document as high, medium, or low. As a result, the queries and documents judged as high-quality are included in the collection. 
During this step, we identify and resolve cases where the documents could provide ambiguous content in terms of their expressed gender stereotype. For instance, to create gender-neutral content when referring to individuals, it is preferred to only use surnames (which are gender-independent) and drop possible first names. For instance, for the query \emph{most important scientists in the world} (item E-3 in the category Cognitive Capabilities), the text of the document \textit{``Get to know the scientists ... 1. Sara Josephine Baker 2. Ben Barres 3. Lauren Esposito''} is replaced with the text \textit{``Get to know the scientists ... 1. Baker 2. Barres 3. Esposito''} to indicate neutral content. Such ambiguous documents are revised and more precisely formulated.
 

Along with evaluating the quality of the queries and documents, the dataset reviewers judge the possible \emph{expected stereotype} of each query. We adopt the definition of expected stereotype from \citet{cuddy2008warmth}: ``The anticipated characteristics and behaviour in males and females, confirming gender stereotype theory of women and men being perceived differently in the social dimensions of warmth and competence''. These gender stereotypes become particularly apparent in the areas such as physical appearance, intelligence, interests, social traits, and occupational orientations~\cite{glick1999sexism}, covered in our categories. To this aim, both data reviewers independently judge the expected stereotype of each query as `toward female', `toward male', and `unspecified/unknown'. The expected stereotype is assigned to the first two options only if a full agreement between the annotators is achieved, otherwise it is set to `unspecified/unknown'. For instance, the expected stereotype for the query \emph{how to easily clean at home} (item F-4 of the Domestic Work category) is set to female. We should highlight that the objective of annotating expected stereotypes for queries is to provide a general guideline regarding the possible direction of stereotypes. As discussed in the next section, this can for instance enable evaluating to what extent a search engine enhances or mitigates such stereotypes. 




\section{Potential Use-Cases and Impacts}\label{sec:usecases}
The proposed dataset enables a variety of research activities concerning both model-related/algorithmic and social aspects of bias and fairness in IR. As these two strands of research are gaining increasing attention from the IR community (and other communities), we expect the \dataset dataset to grow in demand and contemplate various use-cases: 

\paragraph{Search engine behavior analysis} Providing three sets of relevant and non-relevant documents with different gender indications (male, female, and neutral) for every query, \dataset allows to evaluate IR models in terms of their gender bias in retrieval results. The simplest scenario would constitute comparing relevance scores and rankings of documents with different gender indicators provided by an IR model when given a gender-neutral query. For instance, such experiments might show that, for a given category of queries, a retrieval system consistently ranks male-indicated relevant documents higher than equally relevant female-indicated documents. 
Such a finding could provide evidence for a specific gender bias aspect captured or amplified by the IR system~\cite{rekabsaz2020neural,bigdeli2021orthogonality,bigdeli2021exploring,kay2015unequal}. 
This is a harmful bias since it can lead to unfair behaviour of the respective IR system, \ie \emph{the system systematically and unfairly discriminates against certain individuals or groups of individuals in favour of others}~\cite{DBLP:journals/tois/FriedmanN96}.

\paragraph{Studies on gender bias in human judgement} The dataset can be used to conduct user studies on the relevance judgements of users of IR systems. 
It provides a valuable resource and foundation for 
the discovery and analysis of gender biases in more detail. 
For instance, such a study may show that for certain types of queries users deem female-indicated documents to be more relevant than male-indicated or neutral documents, or vice versa. 
Adding different demographics of the study participants as additional variables to the study may uncover interesting aspects about the influence of cultural or socio-demographic traits on users' perception of relevance.
This could help, for instance, to answer questions like: Are users with an Eastern cultural background more or less sensitive to gender-specific content in search results than their Western counterparts? Are female users generally more aware of retrieval contents biased towards maleness?
Does the judgement of gender-specific relevance differ between users of different ages?


\paragraph{Human query generation analysis} Another approach to reveal and study gender biases among users would be to consider 
user-generated queries. A corresponding study could provide users with one of the three gender-indicated versions of a document, and ask them to formulate a query they would use to find the document. 
The resulting user-generated queries can then be analyzed for gender pointers. If users tend to create gender-neutral queries to search for a male-indicated document, it may indicate that the relation of the document's topic to the male gender is implicitly assumed by the users, which may be indicative of another kind of gender bias, \ie a cognitive stereotypical bias.
\\
\\
Findings of the kind sketched above for various use-cases will have an important impact on our understanding of data and algorithmic, but also cognitive, biases in IR research. They will stimulate further research on ways to disclose such biases to users of IR systems, and to mitigate them in an effort to create bias-aware and fairer search engines.


\section{Limitations and Ethical Considerations}\label{sec:limitations}
While the \dataset is, in our opinion, a valuable resource for IR researchers and practitioners studying gender-related biases in search engines, there exist several limitations one should be aware of when using \dataset.

\paragraph{Binary setting of gender}
During the creation of \dataset, we chose to restrict queries to male and female gender indications, neglecting other genders humans may identify themselves with. This was a pragmatic choice, 
not least taken because of the lack of vocabulary in the English language to denote or reflect genders beyond such a binary setting. Still, acknowledging the limitation as this might, to some extent, increase the belief that human beings can be unequivocally  divided into two gender classes~\cite{hyde2019future}.

\paragraph{Western cultural bias}
The original queries, while inspired by an existing gender role framework~\cite{behm2009effects}, were created by one of the authors, who has  
a Western cultural background.
This also holds for the two annotators 
who evaluated the quality of the documents and the  expected gender stereotypes in queries. 
While they are experts in gender studies and therefore particularly sensitive about gender issues, judgements can barely be completely free from subjectivity, especially considering that both creator and annotators are all female. 
In addition, the cultural bias of the annotators might have also impacted the `expected stereotypes', annotations which aimed to mirror the negative  gender-related stereotypes typical of Western culture. Indeed, these `traditional' stereotypes are based often on old-fashioned beliefs, such as assuming a romantic relationship to be heterosexual, which might not be any longer so present in many Western individuals.

\paragraph{Dataset size}
We are well aware that \dataset is a relatively small-scale dataset. However, it is to the best of our knowledge the first dataset of its kind. Its query composition has been based on a well-grounded framework of stereotypical gender role dimensions~\cite{behm2009effects}, 
and it contains meta-annotations of expected gender stereotypes~\cite{cuddy2008warmth}.


\section{Conclusion}\label{sec:conclusions}

We introduced \dataset, an annotated query-document dataset for studying gender representation bias in IR systems and search results.
\dataset comprises 118 bias-sensitive but gender-neutral queries, organized along 7 different topical categories, including physical capabilities, child care, and appearance.
Each query is accompanied by 6 document versions: one set of 3 relevant documents, expressed in neutral, male, and female wording, respectively; one set of 3 non-relevant documents, also expressed in neutral, male, and female wording.
This results in 708 documents related to the 118 queries. We outlined several use cases and discussed the dataset's potential impact, as well as its limitations.
We hope that \dataset represents a valuable resource for many IR researchers and will stimulate further research on the investigation of data, algorithmic, and cognitive gender biases in IR systems, as well as novel technologies to mitigate harmful gender biases and lead to a new generation of gender-fair IR systems.

\begin{acks}
This work received financial support by the Austrian Science Fund (FWF):
P33526 and DFH-23; and by the State of Upper Austria and the Federal Ministry of Education, Science, and Research, through grants LIT-2020-9-SEE-113 and LIT-2021-YOU-215.
\end{acks}

\bibliographystyle{ACM-Reference-Format}
\balance
\bibliography{reference}

\end{document}